\documentclass[sn-mathphys-ay]{sn-jnl}

\usepackage{graphicx}%
\usepackage{multirow}%
\usepackage{amsmath,amssymb,amsfonts}%
\usepackage{amsthm}%
\usepackage{mathrsfs}%
\usepackage[title]{appendix}%
\usepackage{xcolor}%
\usepackage{textcomp}%
\usepackage{manyfoot}%
\usepackage{booktabs}%
\usepackage{algorithm}%
\usepackage{algorithmicx}%
\usepackage{algpseudocode}%
\usepackage{listings}%
\usepackage{xcolor,subfig,lipsum}
\usepackage{natbib}
\usepackage{mathtools}
\usepackage[thinc]{esdiff}

\theoremstyle{thmstyleone}

\theoremstyle{thmstyletwo}%
\theoremstyle{thmstylethree}%

\begin{document}

\title[Article Title]{A Flexible Parametric Approach to Synthetic
Patients Generation Using Health Data}

\author*[1,2,3]{\fnm{Marta} \sur{Cipriani}}\email{marta.cipriani@uniroma1.it}

\author[1]{\fnm{Lorenzo} \sur{Di Rocco}}

\author[3]{\fnm{Maria} \sur{Puopolo}}

\author[1]{\fnm{Marco} \sur{Alfò}}

\affil*[1]{\orgdiv{Department of Statistical Sciences}, \orgname{Sapienza University of Rome}, \orgaddress{\city{Rome}, \country{Italy}}}

\affil[2]{\orgdiv{Mathematical Institute}, \orgname{Leiden University}, \orgaddress{\city{Leiden}, \country{Netherlands}}}

\affil[3]{\orgdiv{Department of Neuroscience}, \orgname{Istituto Superiore di Sanità}, \orgaddress{\city{Rome}, \country{Italy}}}

\abstract{Enhancing reproducibility and data accessibility is essential to scientific research. However, ensuring data privacy while achieving these goals is challenging, especially in the medical field, where sensitive data are often commonplace. One possible solution is to use synthetic data that mimic real-world datasets. This approach may help to streamline therapy evaluation and enable quicker access to innovative treatments.
We propose using a method based on sequential conditional regressions, such as in a fully conditional specification (FCS) approach, along with flexible parametric survival models to accurately replicate covariate patterns and survival times. To make our approach available to a wide audience of users, we have developed user-friendly functions in R and Python to implement it. We also provide an example application to registry data on patients affected by Creutzfeld-Jacob disease. The results show the potentialities of the proposed method in mirroring observed multivariate distributions and survival outcomes.}


\maketitle

\maketitle

\section{Introduction}\label{sec1}

Synthetic data consists of artificially created datasets designed to resemble real-world data. They offer an alternative to original datasets and can be publicly shared, promoting transparency and reproducibility in research. This approach helps address data privacy challenges by replacing observed values with values sampled from probability distributions that maintain the statistical properties of the original data. As a result, synthetic data can be used to create virtual patient cohorts that closely resemble real-world datasets while protecting individual privacy. Moreover, this method overcomes the limitations of traditional anonymization techniques, which have become increasingly vulnerable due to technological advancements \citep{rocher}.

While the use of synthetic data in medical fields such as image analysis, multi-omics data integration, drug discovery, and precision medicine is widespread nowadays \citep{grewal, khanani}, one promising but less explored application is the use of synthetic data in clinical trial design \citep{cipriani}. Synthetic data in clinical trials often requires the generation of synthetic survival variables. This kind of variable presents unique features and requires specialized considerations compared to other types of variables. Generating synthetic survival data involves accurately modelling time-to-event data, which includes handling censored observations, varying follow-up times, and preserving the underlying hazard and survival functions. These tasks are intricate and demand specialized techniques to ensure that the synthetic data not only mirrors the original dataset in its statistical properties but also replicates the complex temporal patterns inherent in survival analysis.

Although the literature on synthetic data generation methods continually expands with increasingly sophisticated approaches, parametric methods for generating synthetic survival variables remain underexplored. Our work addresses this gap by proposing a technique specifically designed for generating synthetic survival data, enhancing the robustness and applicability of synthetic datasets in medical research. This approach may be particularly beneficial in rare diseases, where health data are often limited. Despite the relative availability of global data, individual centers often have insufficient patient data for rare diseases \citep{rollo}. Moreover, in cases where no approved treatment exists, the ability to concentrate patient recruitment into the experimental arm of a trial can be highly advantageous. Synthetic data generation can help create control groups that mirror the characteristics of the target population, leading to more efficient allocation of resources and potentially accelerating the development of new therapies.

Advancements in synthetic data generation have introduced a variety of machine and deep learning methods, such as Classification and Regression Trees (CART; \cite{reiter}), Random Forests (RF; \cite{caiola}), Bayesian Networks (BN; \cite{kaur}), and Conditional Tabular Generative Adversarial Networks (CTGAN; \cite{xu}). Each of these methods has its advantages. CART and RF, for example, are decision-tree-based methods that partition data based on covariates values. However, these models often struggle with complex, high-dimensional datasets. RF, in particular, despite being an ensemble method that improves accuracy, may still lack the flexibility to model survival data accurately \citep{akiya}. BN models, which use probabilistic graphical models to capture dependencies between variables, offer flexibility but are limited by their computational intensity and their reliance on strong assumptions about dependencies \citep{akiya}. CTGANs represent a more recent development of generative adversarial networks (GANs; \cite{gan}), originally developed for generating realistic data such as images. While CTGANs show promise in generating realistic synthetic tabular data, they often require large datasets to train effectively and can suffer from issues such as mode collapse, where the diversity of the original data is not fully captured \citep{akiya}. Moreover, deep learning models like CTGANs are often "black boxes," and this makes it difficult to understand how specific features influence outcomes, which is problematic in clinical research where transparency and understanding the role of covariates are essential for ethical and practical reasons \citep{azizi}.

In contrast, parametric methods can offer a more interpretable and robust approach for generating synthetic data in survival analysis. Traditional parametric models, such as exponential, Weibull, or Gompertz distributions, assume specific distributions for survival times and provide efficient estimates when these assumptions hold true. Flexible parametric methods, such as spline-based models, further extend these benefits by accommodating complex hazard functions and time-dependent effects without sacrificing interpretability \citep{royston}. Additionally, recent studies have highlighted that parametric models can be adapted to various clinical trial scenarios, including generating synthetic data to mirror real-world patient populations for evaluating treatment effects and generalizing findings \citep{rollo}.

From a methodological standpoint, the only comprehensive approach to generating synthetic survival data using parametric methods has been proposed by \cite{smith}. Their work highlights the advantages of using parametric models for this purpose, especially in generating synthetic time-to-event data. However, from a programming perspective, the only available package in R that supports parametric synthetic data generation is \texttt{synthpop} \citep{nowok}. Despite its utility in creating synthetic datasets, \texttt{synthpop} focuses primarily on non-parametric methods, such as decision trees and other machine learning techniques, and does not accommodate the generation of survival data using parametric methods, which limits its scope in the analyzed context. 

To address this gap, we propose in this work a method based on sequential conditional regression and flexible parametric survival models to emulate observed survival data (Section \ref{sec2}). In addition, we introduce user-friendly R and Python functions called \texttt{fleSSy}\footnote{The source code is available at \url{https://github.com/ldirocco/fleSSy}} (FLExible Survival SYnthetic), that allow for seamless implementation of these methods, extending the capabilities of parametric models for synthetic data generation beyond the constraints of existing tools. \\
We demonstrate the potential of this method through an application to registry data on patients affected by Creutzfeldt-Jacob disease in Section \ref{sec3}, and we conclude with a discussion in Section \ref{sec4}.


\section{Methods}\label{sec2}

This section outlines the methodology employed by \texttt{fleSSy} to generate synthetic survival data. We use a flexible parametric proportional-hazards model to fit observed covariates, enabling the capture of complex survival patterns. Following this, we reconstruct the joint distribution of covariates using a fully conditional specification (FCS) approach, ensuring that interdependencies between variables are preserved. We then simulate survival times, accounting for dropout, right-censoring, and death, to create a synthetic dataset that mirrors real-world scenarios. Finally, we evaluate the quality and utility of the synthetic data in comparison to the original dataset.

\subsection{Fitting the survival model on observed covariates}
To generate synthetic survival data, we begin by modelling individual survival times using the flexible parametric proportional-hazards model proposed by \cite{royston}. This model allows for the flexibility needed to capture complex hazard functions. The key idea is to model the logarithm of the baseline cumulative hazard function as a "natural" cubic spline function of log time, which provides a smooth and flexible fit to the data. 

\begin{figure}[h]
    \centering
    \includegraphics[width=1\linewidth]{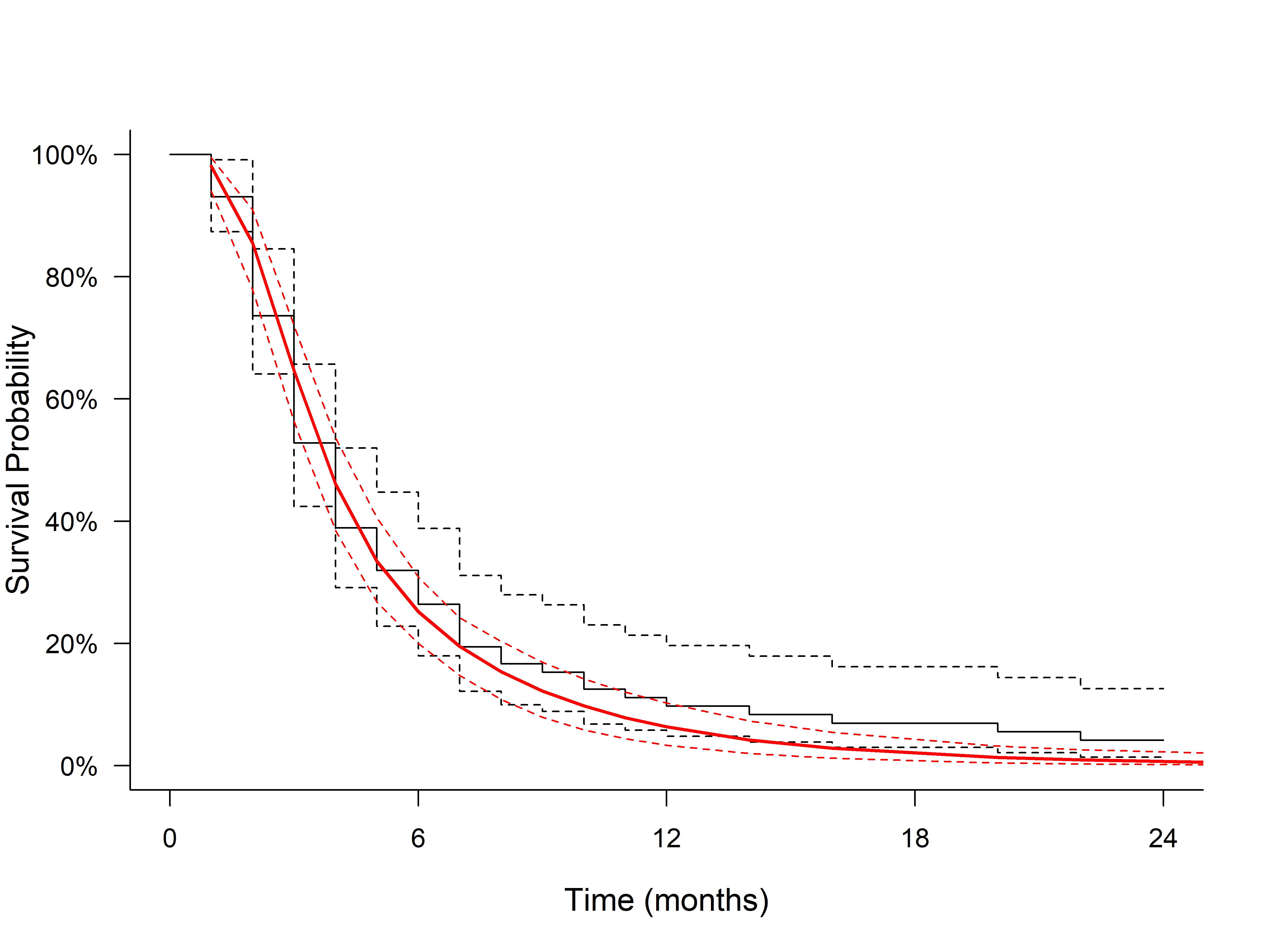}
    \caption{Fitted survival from the saturated PH spline model (red line) against Kaplan-Meier estimates (black line). Dashed lines represent the confidence interval estimates. A single population Kaplan-Meier curve is drawn with covariates set to their mean values in the data; for categorical covariates, the means of the 0/1 indicator variables are used.}
    \label{fig:flex}
\end{figure}

Specifically, the survival function $S(t)$ is transformed using a link function $g(\cdot)$, such that for a proportional-hazards spline model with a fixed covariate vector $\textbf{z}$, we have:
\begin{equation}
    g[S(t;\textbf{z})] = \log [- \log S(t; \textbf{z})] = \log H(t; \textbf{z}) = \log H_{0}(t) + \beta^{T}\textbf{z} = s(x;\gamma) + \beta^{T}\textbf{z}
\end{equation}
where $H(t;\textbf{z})$ is the cumulative hazard function, $H_{0}(t)$ is the baseline cumulative hazard function, and $s(x;\gamma)$ is a natural cubic spline. The latter is defined as a cubic spline constrained to be linear beyond boundary knots $k_{\min}, k_{\max}$. In addition, $m$ distinct internal knots $k_{1} < \dots < k_{m}$ are specified, and a natural cubic spline may be written as 
\begin{equation}
    s(x; \gamma) = \gamma_{0} + \gamma_{1} x + \gamma_{2} \upsilon_{1}(x) + \dots + \gamma_{m+1} \upsilon_{m} (x)
\end{equation}
where the $j$th basis function $\upsilon_{j}(\cdot)$ is defined for $j=1, \dots, m$ as
\begin{equation}
    \upsilon_{j}(x) = (x-k_{j})_{+}^{3} - \lambda_{j} (x-k_{\min})_{+}^{3} - (1 - \lambda_{j}) (x-k_{\max})_{+}^{3}
\end{equation}
with $\lambda_{j} = \frac{k_{\max} - k_{j}}{k_{\max}-k_{\min}}$ and $(x-a)_{+} = \max (0, x-a)$. \\
The complexity of the curve is determined by the number of degrees of freedom (d.f.), which, excluding $\gamma_0$, is equal to $m + 1$. When no internal knots are specified (i.e., $m = 0$ or d.f. = 1), the spline function $s(x; \gamma)$ reduces to a linear form, $s(x; \gamma) = \gamma_0 + \gamma_1x$, yielding a baseline cumulative hazard that aligns with a Weibull distribution. This constraint arises from the limited degrees of freedom, which precludes additional curvature in the baseline hazard, thereby restricting it to distributions that can be expressed in closed form. \\
The positioning of internal knots is a matter of consideration. However, as discussed by \cite{durrleman}, the selection of ‘optimal’ knots does not seem to be crucial for achieving a good fit and may even be counterproductive, as the fitted curve might overly mimic small-scale variations in the data. 
Aligning with the recommendations of \cite{royston}, in \texttt{fleSSy} we positioned default boundary knots at the extreme uncensored log survival times. For the internal knots, we utilized the centile-based positions listed in Table \ref{knots}.
\begin{table}[h]
\caption{\small Internal knots placement.} 
\centering
 \label{knots} 
 \begin{tabular}{lccc}
\hline 
\hline \\[-1.8ex]  
 \textbf{d.f} &  & \textbf{Centiles} & \\
 \hline \\[-1.8ex] 
 2 & 50 & & \\
 3 & 33 & 67 &  \\
 4 & 25 & 50 & 75 \\
\hline 
\hline \\[-1.8ex]  
\end{tabular}
\end{table} \\
However, both the number of knots and their position can be chosen by the function user, based on the data that have to be analyzed. For example, in our case, to fit a saturated model to the data we will further discuss in Section \ref{sec3}, we used 3 internal knots positioned as in Table \ref{knots} and Figure \ref{fig:flex} displays the goodness-of-fit of the estimated model. 

As regards the estimation procedure, let $\ell_{i}(\theta)$ be the likelihood contribution for a generic observation, so that the likelihood over the entire sample is $\prod_{i} \ell_i(\theta)$. Then, the log-likelihood for the $i$th individual can be written as
\begin{equation*}
    \log \ell_{i}(\theta) = \delta_i \biggl[ \log \biggl( s' \bigl(\log(x_i); \gamma \bigr) \biggr) + \eta_i \biggr] - \exp(\eta_i)
\end{equation*}
where $\delta_i$ is the event indicator (i.e., $\delta_i = 1$ if the observation $i$ is uncensored), and $\eta_i = \beta^T \textbf{z}_i$.
To obtain maximum likelihood (ML) estimates, starting values for $\beta$ and $\gamma$ are obtained through least-squares regression of the logarithms of the initial cumulative hazard against $x_i$, $z_i$, and the spline basis functions. Then the ML estimation is performed using optimization functions available in the software. Namely, the \texttt{optim} function in R library \texttt{stats}, and \texttt{optimize} into \texttt{SciPy} for Python.

\subsection{Generation of synthetic covariates distributions}
We employed a FCS approach \citep{fcs} to estimate the joint distribution of the covariates used in the survival model. This method helps us avoid explicitly specifying the joint distribution and it rather proposes to build the full joint distribution using conditional distributions. Let the data be represented by the \(n \times p\) matrix $\textbf{X}$. Let \(X_j\) be the \(j^\mathrm{th}\) column in $\textbf{X}$, and \(X_{-j}\) indicates the complement of \(X_j\), that is, all columns in $\textbf{X}$ except \(X_j\). Then, FCS specifies the multivariate distribution $P(X \vert \theta)$ through a set of conditional densities $P(X_j \vert X_{-j}, \theta_j)$. If the conditional distributions are compatible, which means that their density ratios $P(X_j \vert X_{-j}) / P(X_{-j} \vert X_{j})$ can be factorized into the product of integrable functions $\phi_j(X_j)$ and $\nu_{-j}(X_{-j})$ for all $j$ \citep{besag}, the algorithm is a Gibbs sampler, a Bayesian simulation technique that samples from the conditional distributions to obtain samples from the joint distribution \citep{gelfand, casella}. 

We used an approach similar to the one adopted in the algorithm for multiple imputation by chained equations (MICE; \cite{buren}). In our method, synthetic values for each variable are generated sequentially from their conditional distributions based on the values of previously synthesized variables, with the parameters estimated from the observed data. Users can specify the types of models to use for synthesis, the order in which variables should be synthesized (using the \texttt{seq} argument), and the set of predictors to include in the synthesis model (using the \texttt{pred} argument). Therefore, given the algorithm structure, it is crucial to determine the order in which variables should be synthesized. In conventional applications of the Gibbs sampler, the full conditional distributions are derived from the joint probability distribution \citep{gilks}. In the MICE algorithm as in ours, the conditional distributions are under the direct control of the user, and so the joint distribution is only implicitly known, and may not exist. While the latter is undesirable from a theoretical point of view (since we do not know the joint distribution to which the algorithm converges), in practice it does not seem to hinder useful applications of the method.\\
It is worth noting that, theoretically, different factorizations of the joint distribution can lead to the same overall distribution. However, when estimating parameters from each factorization, the use of different conditional models may lead to different estimates. This potential variation arises because conditional independence assumptions or structural choices made in the factorization can affect the accuracy and consistency of parameter estimates derived from each conditional distribution. 

Unlike MICE, where the algorithm generates synthetic values based on all other columns in the data, by default our approach restricts predictors to only those variables that have already been synthesized. This restriction ensures that only previously synthesized covariates are used as predictors, which may be essential for maintaining data anonymization and integrity in the imputation process. However, it is possible to include certain variables as predictors without synthesizing them. 

Let $\textbf{X}^{(k)}$ be the dataset at iteration $k$, where each variable is synthesized sequentially. The imputation for $X_{j}$ is based only on the variables that have been previously synthesized up to iteration $k$:
\begin{equation}
    X_{j}^{(k)} = g_{j}(\textbf{X}_{-j; 1:k}) + \epsilon_{j}
\end{equation}
where $g_j$ is the regression model for the $j$-th variable, $\textbf{X}_{(1:k)}$ denotes all variables synthesized up to iteration $k-1$ and those synthesized at iteration $k$ before $X_{j}$, and $\epsilon_{j}$ represents the residual error. 

The form of the $g_j$ depends on the variable type. Our function, \texttt{fleSSy}, simplifies this process by offering pre-built functions that automatically select and apply the appropriate regression models based on the type of covariate — whether binary, categorical, or continuous. For example, default methods include a logistic regression model for binary variables, a polytomous logistic regression for factors with more than two levels, and an ordered polytomous logistic regression for ordered factors with more than two levels. Continuous variables are modelled using linear regression with inverse normal-rank transformations to approximate non-normal distributions. \\
The other available regression methods are:
\begin{itemize}
    \item Predictive mean matching
    \item Linear regression
    \item Linear regression with specified prior information
    \item Linear regression with Lasso penalty
    \item Logistic regression with Lasso penalty
    \item Proportional odds model
    \item Linear discriminant analysis
\end{itemize}
It is also possible to write customized regression functions. 

To generate synthetic data which closely mimics the original data, we require two essential information in our initial design matrix: a binary variable indicating the dropout status and the time of entry into the study for each individual. The user can choose which additional covariates to synthesize.

\subsection{Generating simulated survival times}
After generating the complete set of covariates, we simulate survival times using the survival model we have previously fitted to the original data. We consider the date of the end of the study as our right censoring time. 

The subsequent step is to code the survival status by combining dropout, right-censoring, and death. 

It is important to note that we synthesize each individual’s covariate pattern from a distributional range, rather than a specific real-world individual. Therefore, we do not model survival times directly from any real person. Consequently, there is no connection between a synthetic individual and any individual in the real-world dataset.

\subsection{Comparison between the original and synthetic distributions}
We used rigorous statistical methods to examine how well the synthetic cohort matched the original one. To achieve this, we combined the original and synthetic records to measure how accurately the data values predict the source of the records, using an approach based on the propensity score.

Following the recommendation of \cite{snoke}, the most commonly suggested utility measure in this context is the propensity score mean squared error (pMSE) or its standardized ratio. The pMSE is calculated as the mean-squared difference between the predicted probability that a record originates from the synthetic data and the true proportion of records from the synthetic data, namely: \begin{equation*}
    pMSE = \frac{1}{N} \sum_{i=1}^{N} (\hat{p}_i - c)^{2},
\end{equation*}
where $N$ is the number of records in the dataset obtained by merging the original and the synthetic ones, $\hat{p}_i$ is the estimated propensity score for record $i$, and $c$ is the proportion of synthetic data in the merged dataset. \\
The expected value and variance of the null pMSE, as derived by \cite{snoke}, are given by:
\begin{equation}
\label{eq1}
    E(pMSE) = \frac{(M-1)}{8N}
\end{equation}
and
\begin{equation}
\label{eq2}
    Var(pMSE) = \frac{2(M-1)}{(8N)^{2}}
\end{equation}
where $M$ represents the total number of predictors in the model ($M \geq p$).  \\
With the expressions in \ref{eq1} and \ref{eq2} we can define a standardized version of the pMSE.
Building on the findings of \cite{raab}, we suggest a practical threshold for evaluating utility: if all standardized $pMSE$ are below 10, and ideally below 3, additional adjustments may not be required, suggesting that the synthetic data is sufficiently useful. 

Another approach to utility measures involves grouping the original and synthetic data by constructing tables based on their values and computing measures of difference between these tables. We used the Welch two-sample t-test, the Mann-Whitney test and the two-sample test for equality of proportions to evaluate adherence across both continuous and categorical variables. Moreover, we evaluated the differences in the survival outcomes of the virtual and actual cohorts using the Log-rank test.  

Recognizing that a high degree of statistical similarity may not necessarily imply true similarity, especially when considering the intricate interplay between variables, and appreciating the critical importance of these interrelationships, we conducted a thorough stratified survival analysis. This enabled us to gain further insights into the influence of factors on distinct patient subgroups.


\section{Case study}\label{sec3}
The feasibility of the proposed methods of synthetic patient generation is shown through an application to data collected by the National Creutzfeldt-Jakob Disease Surveillance Unit (NCJDSU) at the Istituto Superiore di Sanità (ISS) in Italy, which monitors cases of sporadic Creutzfeldt-Jakob disease (CJD) diagnosed in Italy between 1993 and 2019. \\
CJD is a rare and invariably fatal neurodegenerative disorder, primarily caused by the accumulation and spread of misfolded prion proteins \citep{ladogana}. Approximately 85\% of cases are sporadic (sCJD), occurring randomly across the globe, while genetic and acquired forms are less frequent, with variable incidence rates depending on the country \citep{zerr}. Although the precise cause of sCJD remains unknown, sporadic cases raise public health concerns due to occasional interhuman transmissions, warranting continued surveillance efforts \citep{watson}. \\
The only established risk factors for sCJD are advanced age and homozygosity for methionine (M) or valine (V) at codon 129 of the \emph{PRNP} gene, which codes for prion protein \citep{alperovitch}. sCJD typically affects individuals over 60, with peak incidence occurring in the seventh decade, and is rarely seen in those under 50. The median survival time for individuals with sCJD is approximately 5 months from the onset of symptoms \citep{pocchiari}. The codon 129 polymorphism also influences susceptibility, with around 70\% of cases occurring in individuals with the MM genotype. Furthermore, distinct clinical and neuropathological subtypes have been identified based on variations at codon 129 of \emph{PRNP} and the accumulation of prion protein (PrPSc) in the brain \citep{baiardi}. sCJD manifests as rapidly progressive dementia with motor dysfunction, eventually leading to akinetic mutism and full dependence \citep{puoti, parchi}. \\
The clinical diagnosis of sporadic CJD relies on the presence of dementia along with at least two other symptoms, such as visual impairment, ataxia, pyramidal or extrapyramidal signs, myoclonus, or akinetic mutism. This diagnosis is further supported by instrumental tests, including electroencephalography (EEG) showing periodic sharp wave complexes and brain magnetic resonance imaging (MRI) detecting hyperintensities in the basal ganglia. \\
The diagnostic criteria have been revised over time and now include the detection of the 14-3-3 protein in cerebrospinal fluid (CSF) since 1998, the use of real-time quaking-induced conversion (RT-QuIC) to detect PrPSc in accessible biological fluids and tissues since 2017, as well as the identification of cortical hyperintensities on brain MRI since 2017 \citep{hermann}. These diagnostic advancements have been integral to improving international epidemiological surveillance and data harmonization.

\subsection{Available data}
Following procedures of the Italian NCJDSU, patients' data on family history, clinical characteristics, and laboratory or instrumental findings were collected. Each patient was followed until death or until an alternative diagnosis was confirmed. Only definite (A1) or probable (A2) cases from January 2017 onwards, classified according to the 2017 EU diagnostic criteria \citep{CJDNetwork2021}, were included in the analysis. Patients were selected starting from this date to ensure homogeneity in the sample, given changes in diagnostic tests over time. This selection yielded an initial sample of 421 patients for synthesis. From this sample, we generated a synthetic cohort twice the size ($n = 842$). \\
The covariates synthesized included age at onset, sex, classification (A1 or A2), codon 129 polymorphism, PrPSc glycotype, family history, EEG, detection of the 14-3-3 protein, and MRI findings. Survival was defined as the time between disease onset and death.

\subsection{Results}
The synthetic data closely resembles the covariate patterns found in the original data, ensuring that survival predictions conditional on distinct covariate profiles are appropriately reflected. \\
Table \ref{cens} summarises the synthetic cohort's features and compares them to those of the original population. Notably, the clinic-biological characteristics of the two cohorts did not differ significantly, at least based on the central measures. Specifically, variables such as classification, codon 129 polymorphism, and familiarity show a very strong alignment (p $> 0.99$) between the original and synthetic data, reinforcing the validity of the synthetic cohort in replicating key demographic and clinical features. \\ 
However, further insights may be gained by examining t-distribution-free test statistics on the cumulative distribution function (CDF). 
\begin{table}[h]
\caption{\small Comparison between the original and synthetic CJD cohorts in terms of demographic and clinic-biologic characteristics.} 
\centering
\small
 \label{cens} 
 \begin{tabular*}{\textwidth}{@{\extracolsep\fill}lcccc}
\toprule
\hline 
\hline \\[-1.8ex]  
 \textbf{Characteristic} & \textbf{Original} & \textbf{Synthetic} & \textbf{p-val$^{1}$} & \textbf{S\_pMSE$^{2}$} \\
 & \textbf{(n = 421)} & \textbf{(n = 842)} & & \\
 \hline \\[-1.8ex] 
 \textbf{Classification, n (\%)} & & & $>0.99$ & 0.004 \\
 \quad \textit{A1} & 245 (58.2\%) & 489 (58.1\%) &  & \\
 \quad \textit{A2} & 176 (41.8\%) & 353 (41.9\%) &  & \\
 \textbf{Age at onset, median (Q1, Q3)} & 70 (64, 77) & 70 (64, 77) & 0.82 & 0.218 \\
 \textbf{Sex, n (1\%)} & & & 0.87 & 0.931 \\
 \quad \textit{Male} & 213 (50.6\%) & 430 (51.1\%) & & \\
 \quad \textit{Female} & 208 (49.4\%) & 412 (48.9\%) & & \\
 \textbf{Codon 129, n (\%)} & & & $>0.99$ & 0.069 \\
 \quad \textit{MM} & 165 (61.1\%) & 328 (61.4\%) & & \\
 \quad \textit{MV} & 62 (23.0\%) & 122 (22.8\%) & & \\
 \quad \textit{VV} & 43 (15.9\%) & 84 (15.7\%) & & \\
 \textbf{PrPSc glycotype, n (\%)} & & & 0.97 & 0.605 \\
 \quad \textit{Type 1} & 93 (61.6\%) & 194 (60.6\%) & & \\
 \quad \textit{Type 2A} & 39 (25.8\%) & 86 (26.9\%) & & \\
 \quad \textit{Other} & 19 (12.6\%) & 40 (12.5\%) & & \\
 \textbf{Familiarity, n (\%)} &  &  & $>0.99$ & 0.027 \\
 \quad \textit{Negative} & 297 (97.4\%) & 597 (97.4\%) & & \\
 \quad \textit{Uncertain} & 8 (2.6\%) & 16 (2.6\%) & & \\
 \textbf{EEG, n (\%)} & & & 0.58 & 1.136 \\
 \quad \textit{Atypical} & 221 (54.0\%) & 425 (52.3\%) & & \\
 \quad \textit{Typical} & 188 (46.0\%) & 387 (47.7\%) & & \\
 \textbf{14-3-3 in the CSF, n (\%)} & & & 0.85 & 0.143 \\
 \quad \textit{Negative/Weakly positive} & 85 (25.6\%) & 175 (26.2\%) & & \\
 \quad \textit{Positive} & 247 (74.4\%) & 494 (73.8\%) & & \\
 \textbf{MRI} & & & 0.97 & 0.101 \\
 \quad \textit{No/NA} & 42 (10.0\%) & 83 (9.9\&) & & \\
 \quad \textit{Yes, neg} & 75 (17.8\%) & 155 (18.4\%) & & \\
 \quad \textit{Yes, pos} & 304 (72.2\%) & 604 (71.7\%) & & \\
 \textbf{Survival status, n (\%)} & & & 0.57 & 0.968 \\
 \quad \textit{Alive/Censored} & 15 (3.6\%) & 25 (3.0\%) & & \\
 \quad \textit{Dead} & 406 (96.4\%) & 817 (97.0\%) & & \\
\hline 
\hline \\[-1.8ex]  
\end{tabular*}
\footnotetext[1]{Welch Two-Sample t-test; Mann-Whitney test; Two-sample test for equality of proportions}
\footnotetext[2]{Standardised propensity score Mean Square Error}
\end{table}    \\
Remarkably, for every synthesized variable, the associated standardized pMSE (S\_pMSE) falls well below the suggested threshold of 3, confirming that the synthesized data closely replicates the original covariate structures. For instance, the S-pMSE for all variables, except EEG, is lower than 1, indicating a strong correspondence between the two datasets. 

\begin{figure}[h]
    \centering
    \includegraphics[width=1\linewidth]{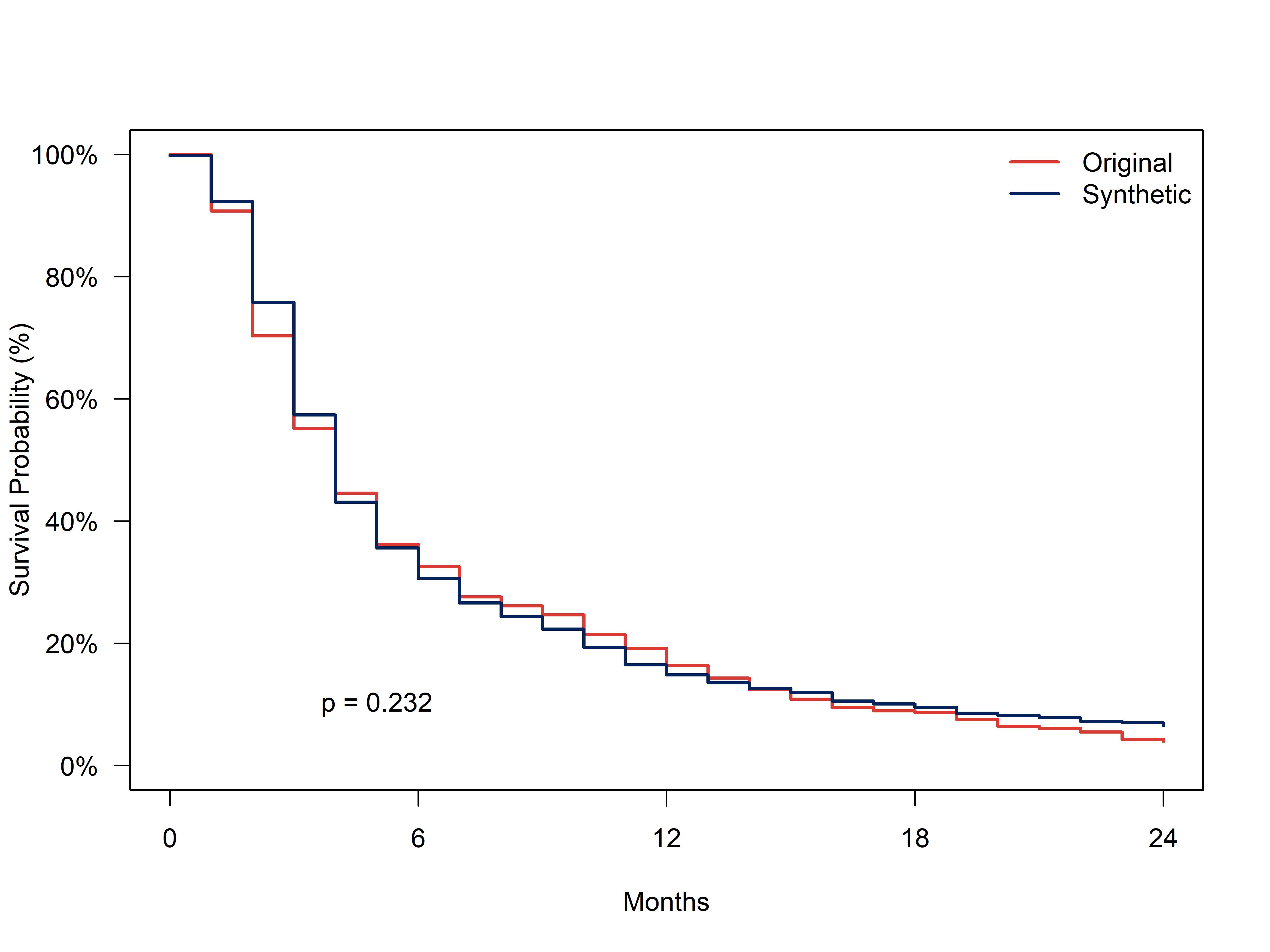}
    \caption{Marginal Overall Survival estimates in the synthetic and original cohorts}
    \label{fig:os}
\end{figure}

When examining survival outcomes, the synthetic data also shows excellent agreement with the original data. The overall survival (OS) curves (Figure \ref{fig:os}) exhibit striking similarity, with a p-value of 0.23 when comparing original and synthesised survival curves through a log-rank test. This suggests that the synthetic data is highly effective in replicating the survival patterns of the original cohort. 

\begin{figure}[h]
 \centering
  \subfloat[OS by age at onset]{\includegraphics[width=.5\linewidth]{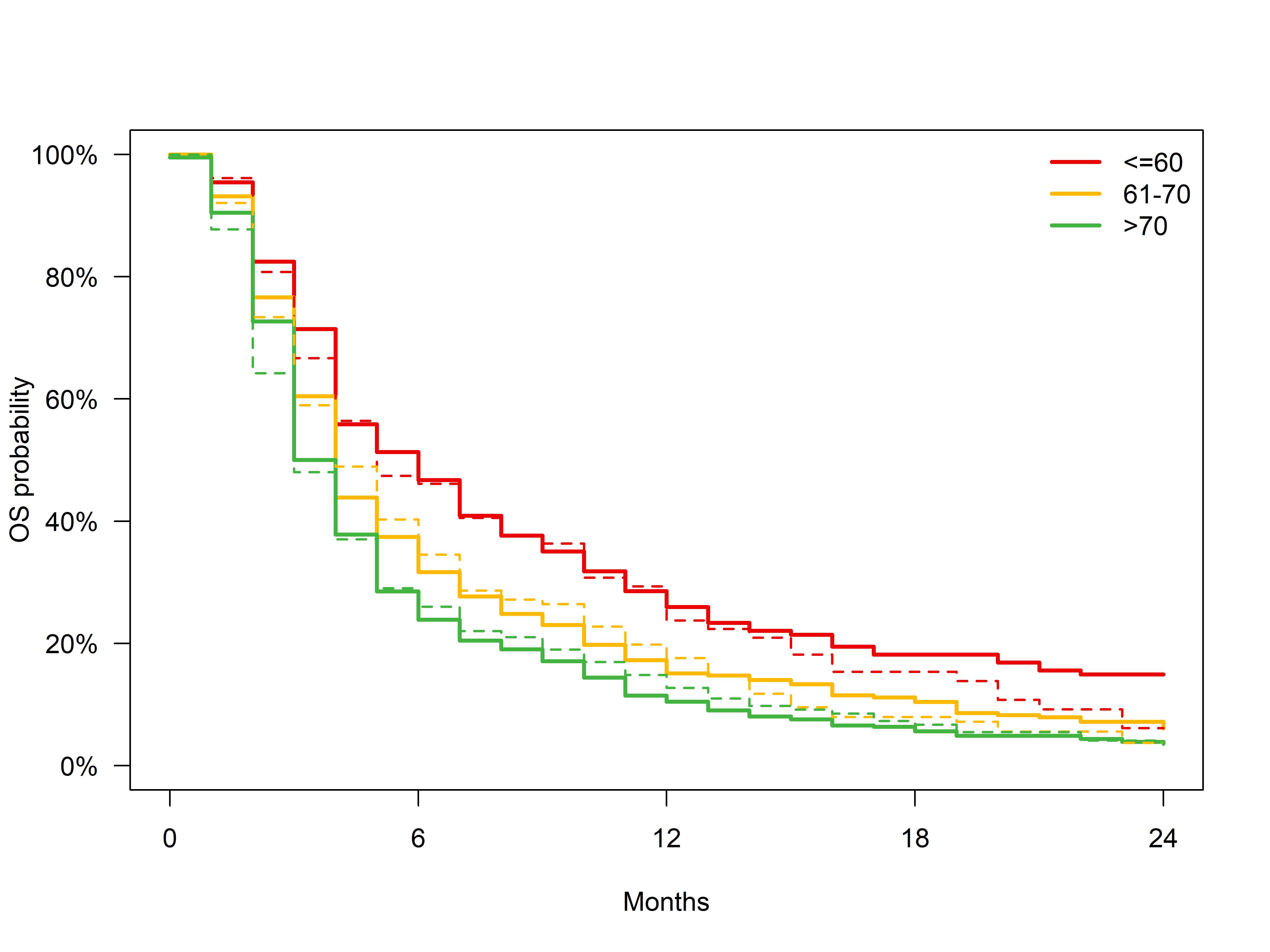}}
  \subfloat[OS by codon 129]{\includegraphics[width=.5\linewidth]{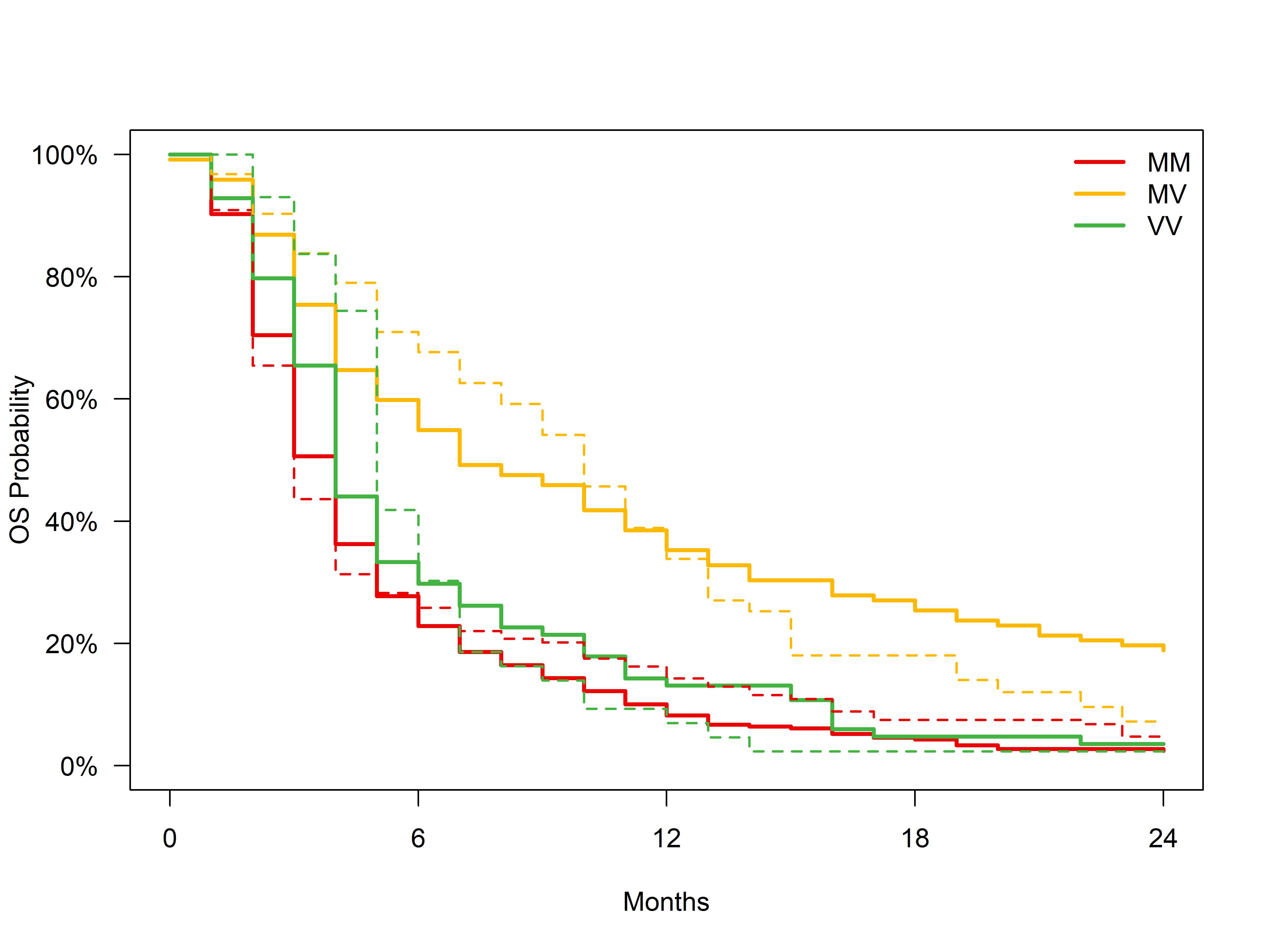}} \\
  \subfloat[OS by PrPSc glycotype]{\includegraphics[width=.5\linewidth]{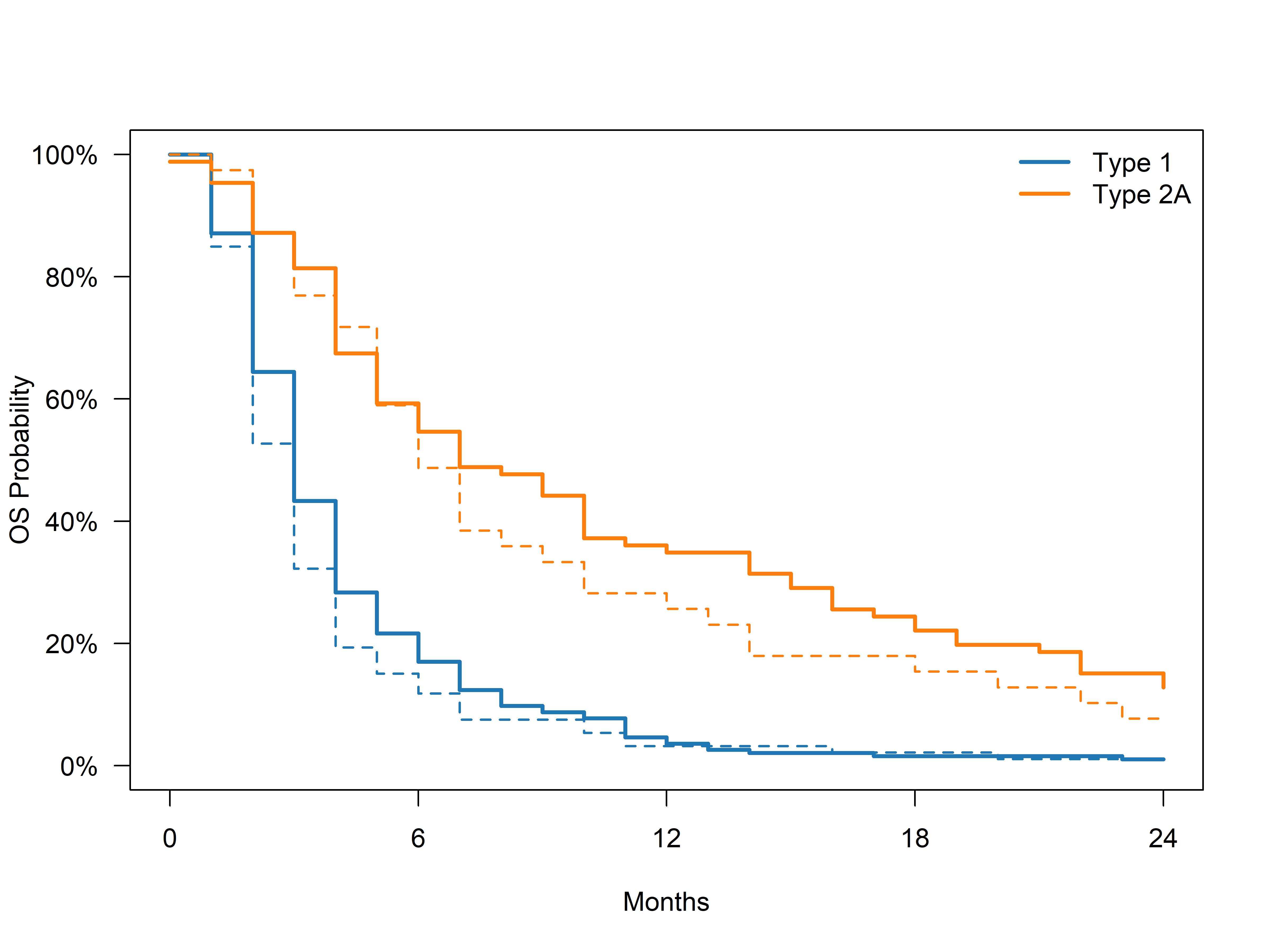}}
  \subfloat[OS by EEG Status]{\includegraphics[width=.5\linewidth]{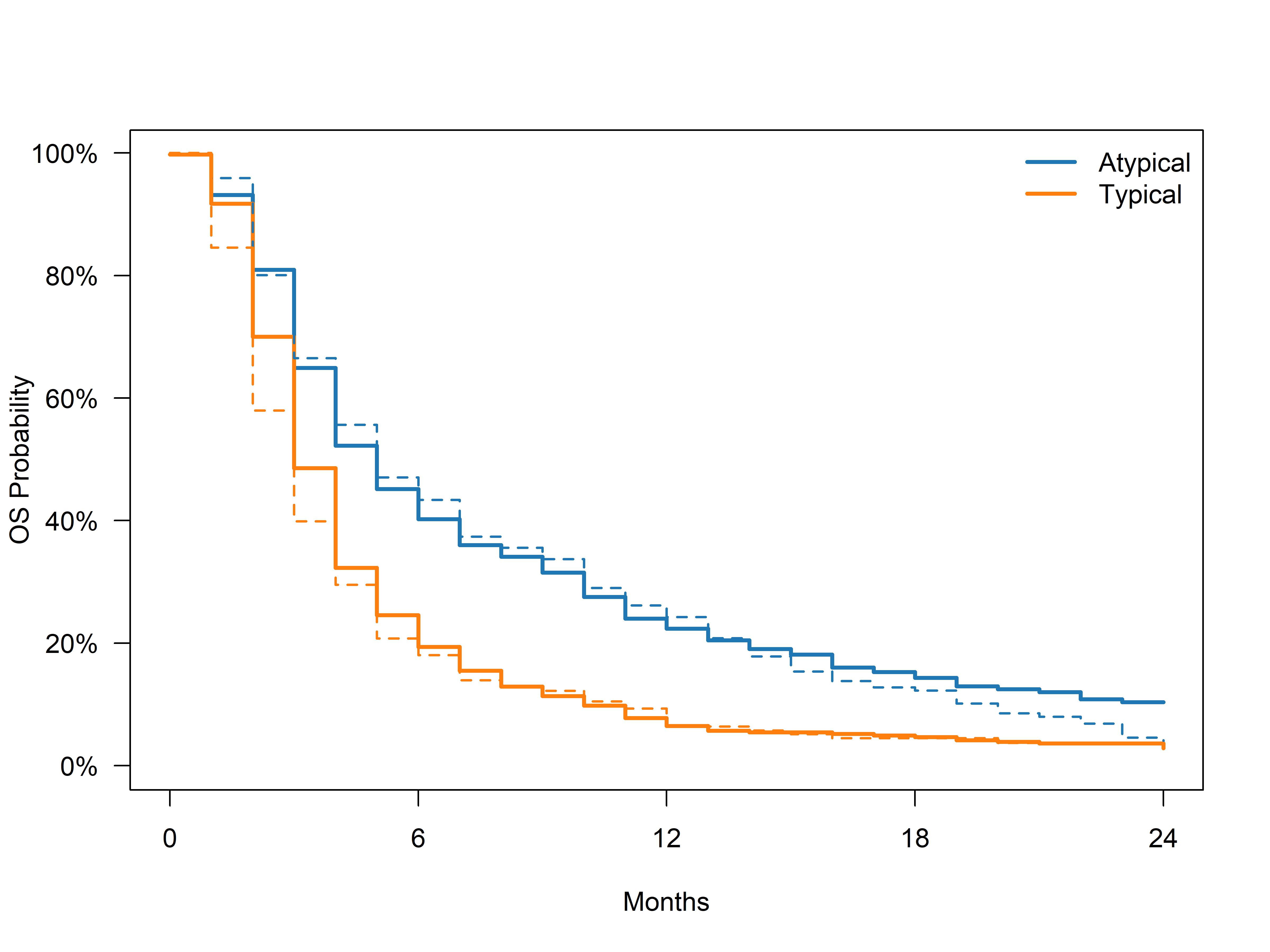}}
  \caption{\textbf{Survival estimates of the original and synthetic cohorts.} Dashed lines refer to the original cohort and continuous lines to the synthetic patients.}
  \label{fig:strat}
\end{figure}

Further stratification of the OS curves by age class, codon 129 polymorphism, PrPSc glycotype, and EEG results (Figure \ref{fig:strat}) reaffirms this consistency. In all stratifications, the synthetic data (represented by continuous lines) closely follows the trends of the original data (dashed lines), showcasing the robustness of the synthetic dataset in capturing survival dynamics across various subgroups. This alignment is further supported by the log-rank test p-values across these stratifications (Table \ref{tab:pvalue}), indicating no significant differences between original and synthetic data across subgroups.

\begin{table}[h!]
    \centering
    \caption{P-value of the log-rank test comparing original and synthetic data for stratification by age, polymorphism, glycotype, and EEG}
    \begin{tabular}{@{}lccc|ccc@{}}
        \toprule
        & \multicolumn{3}{c}{\textbf{Age Class}} & \multicolumn{3}{c}{\textbf{Polymorphism}} \\ 
        \cmidrule(lr){2-4} \cmidrule(lr){5-7}
        & $\leq 60$ & $61$--$70$ & $>70$ & MM & MV & VV \\ 
        \midrule
        \textbf{P-value} & 0.166 & 0.551 & 0.708 & 0.704 & 0.283 & 0.871 \\ 
        \midrule
        & \multicolumn{2}{c}{\textbf{Glycotype}} & & \multicolumn{2}{c}{\textbf{EEG}} & \\ 
        \cmidrule(lr){2-3} \cmidrule(lr){5-6}
        & Type 1 & Type 2A & & Atypical & Typical & \\ 
        \textbf{P-value} & 0.142 & 0.139 & & 0.265 & 0.131 & \\
        \bottomrule
    \end{tabular}
    \label{tab:pvalue}
\end{table}


\section{Discussion}\label{sec4}
This study illustrates the potential of synthetic patient data, especially in the medical field where privacy concerns and limited access to real-world data often pose significant challenges. The generation of synthetic datasets offers a robust solution to mitigate privacy issues while preserving the statistical features of the original data. This not only enables broader data accessibility but also enhances reproducibility — a critical concern in scientific research. Synthetic data generation holds particular value in clinical trials, where it can provide a viable alternative to real-world datasets, especially when patient recruitment is limited or data-sharing is constrained by ethical or legal concerns, paving the way for the so-called \textit{in-silico} trials \citep{pappalardo}.

Multiple approaches can be employed to generate synthetic data. Deep learning techniques are particularly effective at capturing intricate relationships within data. Despite their ability to produce highly realistic synthetic data, these methods have notable drawbacks. A key limitation is the need for extensive data to train the models effectively, which can be challenging in clinical settings where datasets may be small or incomplete.

On the other hand, parametric models provide a suitable approach for generating synthetic survival data, especially when the dataset size is limited. These methods, which include flexible parametric models like the ones used in this study, are robust, interpretable, and adaptable to various data contexts (e.g. ranging from clinical trials data to registry data), making them ideal for smaller and more unbalanced datasets. Our use of flexible parametric survival models effectively addressed the complexities of survival data, such as time-to-event outcomes and censored observations, while maintaining the essential features of the original dataset. Although in our application the censoring rate was very low (less than 4\%) and no dropout events occurred, our function can account for both aspects. For right censoring, it synthesizes an entry time for each observation in a hypothetical study and then censors all times exceeding the observation period endpoint. As for dropout, this is synthesized as a dichotomous variable directly based on the data.

The results of this study further reinforce the applicability of synthetic patient data. Our synthetic cohort demonstrated minimal differences from the original data, underscoring the accuracy of the proposed method. Importantly, survival outcomes were also highly consistent between the synthetic and original data, with no significant differences in overall survival estimates. This suggests that synthetic data can reliably replicate survival patterns.

Moreover, synthetic data provides an optimal solution to address the issue of small sample sizes, which is a frequent limitation in clinical trials, especially for rare diseases. For instance, in the case of rare conditions such as Creutzfeldt-Jakob Disease (CJD), patient data is often sparse, making it difficult to form control groups or conduct robust analyses. By generating synthetic cohorts, we can increase the sample size and correct potential imbalances in the dataset. However, it is essential to recognize that augmenting the sample size with synthetic data may introduce greater variability in the estimates, as synthetic data lacks the natural heterogeneity of real-world data. This can lead to an overestimation of precision and reduced generalizability of the findings, as the synthetic data reflects model assumptions rather than true population characteristics. Careful consideration of these limitations is essential for accurately interpreting results and ensuring that synthetic data complements rather than replaces empirical evidence. 

Furthermore, synthetic data becomes particularly valuable when dealing with orphan diseases, where there is no approved treatment. In cases where a promising treatment is identified, a synthetic cohort could serve as a control group to test the treatment’s efficacy, thereby avoiding the need to expose patients to control conditions that are known to be less effective. This not only protects patients but also accelerates the process of evaluating new treatments, allowing resources to be focused on the experimental therapy.

However, while synthetic data offers numerous advantages, it also presents several challenges, particularly on the regulatory front. In Europe, regulatory progress is underway with the recent approval of the AI Act, which aims to create a unified legal framework for artificial intelligence applications, including synthetic data. However, there remains a deficiency in specific national regulations concerning the use of synthetic data in clinical environments. Until clear legal guidelines won't be in place, the broader adoption of synthetic data in regulatory submissions may face delays.

Another significant challenge relates to the potential limitations of synthetic cohorts. Synthetic datasets, while designed to mirror real-world data, may not adequately capture unknown confounding factors present in the original data. This limitation means that synthetic cohorts may only be suitable for specific contexts where the data generation process is well understood and controlled. If these unknown confounders are not properly accounted for, the validity of synthetic data in representing real-world outcomes could be compromised, limiting its broader applicability.

Finally, the utility of synthetic data is directly linked to the method used to factorize the likelihood during generation. In our case, we applied a fully conditional specification (FCS) approach, which works well for modeling the joint distribution of covariates through sequential regressions. However, this method assumes conditional independence among variables given the others, which might oversimplify the relationships between covariates. This could limit the applicability of the synthetic data in more complex scenarios. Future work could explore the use of copulas as an alternative method to capture the joint distribution of covariates. Copulas offer a way to model the dependence structure between variables without making the strong assumptions inherent in FCS. By incorporating copulas, we could better account for the intercorrelation between covariates, potentially making the synthetic data more realistic and broadly applicable. \\
Further developments of this work also involve incorporating the fleSSy function into
comprehensive libraries in R and Python for synthetic survival data generation.

\backmatter

\bmhead{Acknowledgements}
We would like to acknowledge the numerous Italian neurologists and patients’ families for allowing the collection of clinical, laboratory, and instrumental CJD data. Special thanks go to Dorina Triple and Luana Vaianella, as well as to Anna Ladogana, responsible for the Italian CJD Registry, for their contributions in sharing and curating the dataset for analyses.

\section*{Declarations}

\bmhead{Competing Interests}
The authors have no competing interests to declare that are relevant to the content of this article.

\bmhead{Funding}
The first author's work has been supported by the Italian Institute of Public Health (Istituto Superiore di Sanità) under a PhD grant.

\bmhead{Materials and Code availability}
The data that support the findings of this study are not openly available due to sensitivity reasons. Data are located in controlled access data storage at the National Creutzfeldt-Jakob Disease Surveillance Unit (NCJDSU) of the Istituto Superiore di Sanità (ISS) in Rome, Italy. The source code is available at \url{https://github.com/ldirocco/fleSSy}.








%
%
\bibliography{article}

\end{document}